\documentclass[a4paper,url]{ETHpaper}
\usepackage[utf8x]{inputenc}
\usepackage{amsmath}
\usepackage{amsfonts}
\usepackage{amssymb}
\usepackage[]{algorithm2e}
\usepackage{graphicx}
\usepackage{url}
\title{Social signals and algorithmic trading of Bitcoin} 
\titlealternative{Royal Society Open Science 2:150288 (2015) \url{http://rsos.royalsocietypublishing.org/content/2/9/150288} }
\author{David Garcia, Frank Schweitzer}
\date{today}
\address{Chair of Systems Design, ETH Zurich\\
Weinbergstrasse 56/58, 8092 Zurich, Switzerland}

\begin{document}
\maketitle
\begin{center}
\today\\
Published in Royal Society Open Science 2:150288 (2015)\\
\url{http://rsos.royalsocietypublishing.org/content/2/9/150288}
\end{center}

\begin{abstract}
  
The availability of data on digital traces is growing to unprecedented sizes,
but inferring actionable knowledge from  large-scale data is far from being
trivial.  This is especially important for computational finance,  where
digital traces of human behavior offer a great potential  to drive trading
strategies. We contribute to this by providing a consistent  approach that
integrates various datasources  in the design of algorithmic traders. This
allows us to derive insights into the principles behind the profitability of
our trading strategies. We illustrate our approach through  the analysis of
Bitcoin, a cryptocurrency known for its large price fluctuations. In our
analysis, we include economic signals of volume and price of exchange for USD,
adoption of the Bitcoin technology, and transaction volume of Bitcoin. We add
social signals related to information search, word of mouth volume, emotional
valence, and opinion polarization as expressed in tweets related to Bitcoin
for more than 3 years. Our analysis reveals that  increases in opinion
polarization and exchange volume precede rising Bitcoin prices, and that
emotional valence precedes opinion polarization and rising exchange volumes.
We apply these insights to design algorithmic trading strategies for Bitcoin,
reaching  very high profits in less than a year. We verify this high
profitability with robust statistical methods that take into account risk and
trading costs, confirming the long-standing hypothesis that trading based
social media sentiment has the potential to yield positive returns on investment.

\end{abstract}

\section{Introduction}

Our online society generates data on the digital traces of human behavior at
unprecedented scales and resolutions. This produces a \emph{data deluge}, in
which researchers are confronted with a vast amount of observational data that
is not the product of carefully designed experiments \cite{Lazer2009}. One of
the main challenges of the scientific community is to develop methods to
extract meaningful knowledge from that data beyond mere descriptive analyses
\cite{Vespignani2012}. This is particularly important in financial trading:
Data can be available to all financial agents, but it is the analysis and its
applications what makes a difference. Within computational finance, the field
of algorithmic trading \cite{Treleaven2013} deals with the implementation and
evaluation of automatic trading strategies, which are often kept in private
companies and away from publicly accessible research. The most common kind of
algorithmic trading is based on the principles of \emph{technical analysis}
\cite{Park2004}, using the time series of prices to formulate predictions
about returns. Technical analysis is often insufficient to derive satisfactory
returns \cite{Biondo2013}, motivating the inclusion of large-scale social
signals and the evaluation through  data-driven simulations on historical
data, called \emph{backtesting} \cite{Preis2013,Curme2014}.  In this article,
we present a set of methods to derive stylized facts from the analysis of
multidimensional economic and social signals, and to apply that knowledge in
the design and evaluation of algorithmic trading strategies. We illustrate an
application of our approach to algorithmic trading of the Bitcoin
cryptocurrency, using a wide variety of digital traces about economic and
social aspects of the Bitcoin ecosystem.

Bitcoin (BTC) is a digital currency designed to operate in a distributed
system without any central authority, based on a cryptographic protocol that
does not require a trusted third party \cite{Cusumano2014}. Introduced in a
2008 paper written under the pseudonym of Satoshi Nakamoto
\cite{Nakamoto2008}, Bitcoin  serves as a technology to transfer money quickly
for negligible fees \cite{VanAlstyne2014}.   One of the first markets to adopt
Bitcoin was the \emph{Silk Road}, a website where illegal commerce became
possible thanks to the relative anonymity of Bitcoin \cite{Christin2013}, in
line with the evidence in search trends that relates Bitcoin usage to computer
expertise and illegal activities \cite{Yelowitz2015}. Since then, the use of
Bitcoin has widely expanded beyond criminal activities: At the time of
writing, Bitcoin is accepted by many legal merchants and charities
\cite{Bitpay2014}, including large businesses like Dell \cite{Dell2014}.
Bitcoin-accepting businesses, exchange markets, and wallet services compose
the \emph{Bitcoin ecosystem} \cite{Cusumano2014}, where different kinds of
agents interact, trade, and communicate through digital channels.  The
increasing adoption of Bitcoin and its online nature  allow us to
simultaneously monitor its  social and economic aspects. Every purchase of
goods or services in Bitcoin leaves  a trace in  a public ledger called the
\emph{Block Chain}, creating a publicly accessible economic network
\cite{Schweitzer2009}. Bitcoin's delocalized technology aligns with the online
interaction of its users through social networks and forums, motivating its
adoption by new users through word-of-mouth  \cite{Garcia2014}. Previous
research has shown how search trends and Wikipedia views  are related to price
changes \cite{kristoufek2013} and to the speculative and monetary aspects of
Bitcoin \cite{Kristoufek2015}, leading to dynamics that combine  search
interest, user adoption, word-of-mouth, and prices \cite{Garcia2014}.

\paragraph{Contributions of this article.}

Based on established principles of time series analysis and financial trading,
we present a framework to derive general knowledge from multidimensional data
on social and economic  aspects of a market.  We apply a general statistical
model to detect temporal patterns in the co-movement of price and other
signals.  Those patterns are tested through a method robust to the empirical
properties of the analyzed data, formulating concise principles on which
signals precede market movements. We combine those principles to produce
tractable trading strategies, which we evaluate over a leave-out sample of the
data, quantifying their profitability.  Our approach, rather than focusing on
improving a particular method, takes a multidisciplinary stance in which we
combine principles from social psychology and economics with methods from
information retrieval, time series analysis, and computational finance.

We apply our framework to the Bitcoin ecosystem, monitoring the digital traces
of Bitcoin users with daily resolution. We combine \emph{economic signals}
related to market growth, trading volume, and use of Bitcoin as means of
exchange, with \emph{social signals} including search volumes, word-of-mouth
levels, emotional valence, and opinion polarization about Bitcoin. Our results
reveal which signals precede changes of Bitcoin prices, a knowledge that we
use to design algorithmic trading strategies.  We evaluate the power of our
strategies through backtesting data-driven simulations, comparing returns with
technical analysis strategies. As a consequence, we test the hypothesis that
social media sentiment predicts financial returns in the Bitcoin ecosystem.

\paragraph{Social signals in finance.}  Understanding the role of social
signals in finance not only has the potential to generate significant profits,
but also has scientific relevance as a research question \cite{Schoen2013}.
Two different research approaches give insights to this question: One is the
\emph{statistical analysis} of social and financial signals in order to test
the existence of temporal correlations that lead financial markets. The second
one applies these signals in \emph{prediction scenarios}, measuring their
accuracy as a validation of the underlying behavior of the system, but not
necessarily of their profitability. The statistical analysis of search engine
data reveals that search trends can predict trading volumes of individual
stocks \cite{Bordino2012}. In addition, stock prices in S\&P 500 are
correlated with tweet volumes \cite{Mao2012}, but the  applicability of these
patterns into trading strategies is yet to be evaluated.

Sentiment in social media is closely related to socio-economic phenomena,
including public opinion  \cite{Gonzalez2010}. This motivates the application
of sentiment indicators in the statistical analysis of financial data. Early
works on the sentiment in specialized forums gave negative results about
their impact on returns \cite{Tumarkin2001}. Further research showed that
emotions in private instant messaging between workers of a trading company
precede stages of market volatility \cite{Saavedra2011}.  The expression of
anxiety in publicly accessible data from general blogs precedes trading peaks
and price drops in the S\&P 500 \cite{Gilbert2010}, and  sentiment in Twitter
can be used to predict movements in large-scale  stock indices
\cite{Bollen2011}. It is important to note that, to date,  there is no
evidence that   such sentiment-based  predictions produce significant returns
on investment \cite{Schoen2013}.

\paragraph{Online Polarization.} While most of previous works on sentiment in
financial markets focus on dimensions of valence or mood, the collective
phenomenon of polarization of opinions is often overlooked. The emergence of
polarization in a society gives early warnings on political and economic
phenomena: Polarization in  social networks of  Swiss politicians precedes
controversial elections \cite{Garcia2014b}, and polarization patterns in the
Eurovision Song Contest appear before states of distrust in the European
economy \cite{Garcia2013b}. With respect to financial markets, speculation
theories point to the role of diverse beliefs in financial transactions
\cite{Hirshleifer1977}, leading to the hypothesis that polarization and
disagreement influence trading volumes and prices \cite{Harris1993}. In this
line, the empirical analysis of polarization in stock message boards shows
that states of disagreement lead to increased volatility \cite{Antweiler2004}.

\paragraph{The missing link.}   To date, there is a significant knowledge gap
between the analysis and application of social signals to trading scenarios.
Findings from statistical analyses alone  are not guaranteed to lead to
profitable strategies at all \cite{Gilbert2010}.  For example,  movements of
the Dow Jones Industrial Average  (DJIA)  can be predicted with mass media
sentiment \cite{Tetlock2007} and Twitter mood \cite{Bollen2011},  but to date
no research has shown that such prediction methods can be profitable in
trading scenarios. Similarly, the analysis of discussion patterns in
specialized blogs predict returns of some technology companies
\cite{DeChoudhury2008}, but it is still open to evaluate the potential returns
of such a predictor. The application of methods that process arbitrarily large
datasets lead to results difficult to apply, for example the predicting power
of search volumes of the query \emph{"moon patrol"} \cite{Challet2013} in
backtesting over the DJIA \cite{Preis2013}. Furthermore, analyses of Twitter
discussions about companies can be applied in a portfolio strategy, yet its
evaluation through backtesting leads to very moderate returns and their
statistical significance is not assessed \cite{Ruiz2012}. In addition, no
previous research has proposed a prediction technique that derives significant
returns on investment from online sentiment data \cite{Schoen2013}. Our
research aims at  closing the gap between these lines of research. To do so,
we unify the statistical analysis and its application to design and evaluate
trading strategies, based on tractable principles with potential impact in the
finance community.

\section{Trading strategy framework}
\label{sec:Framework}
\begin{figure}[ht]
\centering
\includegraphics[width=0.98\textwidth]{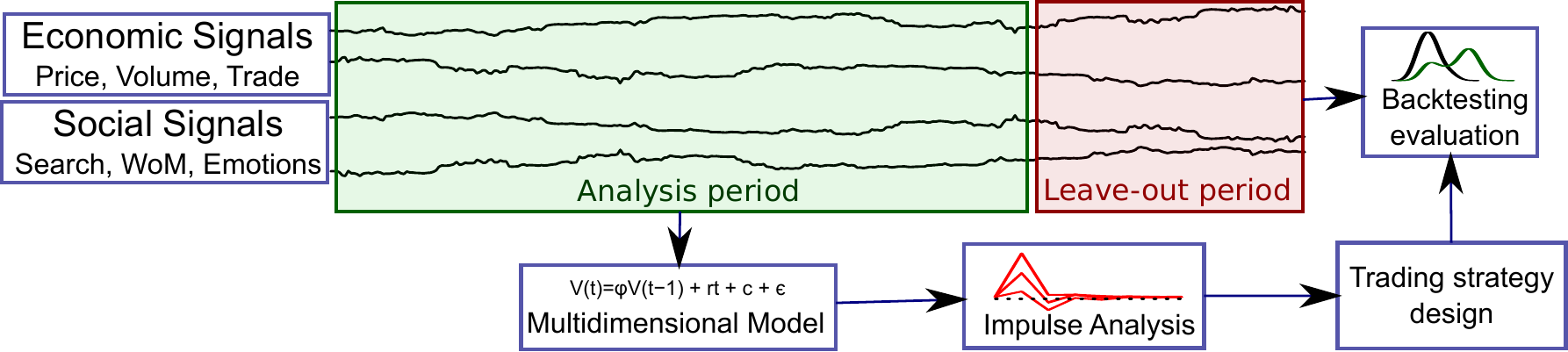}
\caption{\textbf{Framework for analysis of social and economic signals and trading strategy design and evaluation.}}
  \label{fig:Framework}
\end{figure}

To design and evaluate trading strategies, we present a framework that uses a
set of economic and social signals related to the agents of the market under
scrutiny. Among those signals, the only required one is an economic signal of
prices of an asset, namely a stock, currency, or tradable index.  To
understand profitability, we convert the price time series $P(t)$ into a
return time series:
\begin{equation}  
Ret(t)= \frac{P(t)-P(t-1)}{P(t-1)}  
\end{equation} 
which quantifies proportional changes in  the price at every time step. The
data on these signals is divided in an analysis period and a leave-out period, as
depicted in Figure \ref{fig:Framework}. The division in these periods needs to
allocate enough data in the leave-out sample to provide the testing power to
assess the statistical significance of strategy profits. For daily trading,  a
leave-out period of about one year is usually sufficient, but this ultimately
depends on the expected profitability and variance of the trading strategies.

\paragraph{Multidimensional analysis.}  The first step in our framework focuses
on the analysis period, applying a multidimensional model of Vector Auto-Regression 
(VAR) \cite{Whittle1953}, which is commonly used in the analysis of
multidimensional time series in finance
\cite{Tumarkin2001,Adamic2009,Garcia2014}. A VAR models multidimensional
linear relations with given lags, which in our analysis we set to one day.
Thus, given the vector of signals $V(t)$ we fit the equation \begin{equation}
V(t) = \phi V(t-1) + r*t + c + \epsilon \label{eq:VAR} \end{equation} where
$\phi$ is a matrix of weights of the linear relations between variables, $r$
is a deterministic trend vector, $c$ is the vector of constant intercepts, and
$\epsilon$ is a vector of uncorrelated errors. While more advanced models can
be considered, including longer lags and  non-linear terms, we choose the VAR
model of lag 1 for its general character and its proved power to reveal
patterns in finance \cite{Tumarkin2001,Garcia2014}. More complex models might
have higher power to reveal nuance patterns, but at the expense of a loss of
generality  due to the focus on particular systems.

We include all the time series in a single model to avoid the false positives
associated with pairwise Granger tests.  To ensure the correct application of
the VAR model, we need to verify that our analysis is consistent with its
fundamental assumptions:  i) that the elements of $V(t)$  do not have a unit
root, and  ii) that the error term $\epsilon$ has no temporal nor structural
correlations.  We verify the first set of assumptions on the properties of
$V(t)$ by applying a set of tests and transformations prior to the application
of the VAR model. We ensure that our conclusions are robust to the second set
of assumptions by correcting for correlations in the noise term, as explained
in the Materials and methods section.

\paragraph{Impulse analysis.} 

The VAR weights $\phi$  are only informative when there are no  correlations
in the error term $\epsilon$ of equation \ref{eq:VAR}, which is usually not
the case in practice. To extract stylized facts that can be used in the design
of trading strategies, we perform an impulse analysis by measuring  Impulse
Response Functions (IRF) \cite{Lutkepohl2007} while correcting for
correlations in the empirical error. This method simulates the system dynamics
when it receives a shock in one of the variables, applying the VAR dynamics of
Equation \ref{eq:VAR} to  reproduce the changes in the rest of the variables
through time. By recording the changes in each variable, we
can estimate the total size and the timespan of the perturbation produced by
the shock. In essence, the IRF method creates a computational
equivalent of the system under scrutiny, to test its reaction to exogenous
impulses in each of its elements.

\paragraph{Trading strategy design and evaluation.}

The output of the impulse analysis step, shown in Figure \ref{fig:Framework},
is a set of  patterns of Granger-type "causation", i.e. it tests the null
hypothesis of the absence of temporal correlations among the variables. We use
these patterns as stylized facts that indicate which variables precede changes
in price returns. For example, if variable $Y(t)$ has a significant impact on
$Ret(t)$ in the impulse analysis, we will include $Y(t)$ in our trading
strategy design with sign $s_Y$, which takes the value $1$ if the response of
$Ret(t)$ to $Y(t)$ was positive, and $-1$ otherwise. Thus, a predictor based
on $Y(t)$ would be
\begin{equation} 
sign(Ret(t+1)) = sign( s_Y *(Y(t) - Y(t-1)))
\label{eq:prediction} \end{equation}
This way, we predict increases (decreases) in price between time $t$ and $t+1$
if signals with positive responses increase (decrease) between time $t-1$ and
$t$, and vice versa for signals with negative responses.  Since our
multidimensional analysis is robust to confounds between multiple time series, the
findings of impulse analysis can be integrated in  a \emph{Combined} strategy
based on a voting mechanism.  The \emph{Combined}
strategy applies the other predictors and formulates a prediction
corresponding to the sign of the sum of their outputs, i.e. the majority vote.

We evaluate the profitability of the designed strategies  in comparison to the
benchmark of standard strategies, based on the backtesting over the leave-out
sample as indicated in Figure \ref{fig:Framework}. For each strategy, we make
a data-driven simulation of a trader following that strategy, and we record
the profits of that trader on a daily basis. Details on the computational
simulation of financial traders can be found in the Materials and methods
section.

\begin{figure}[ht]
\centering
\includegraphics[width=0.98\textwidth]{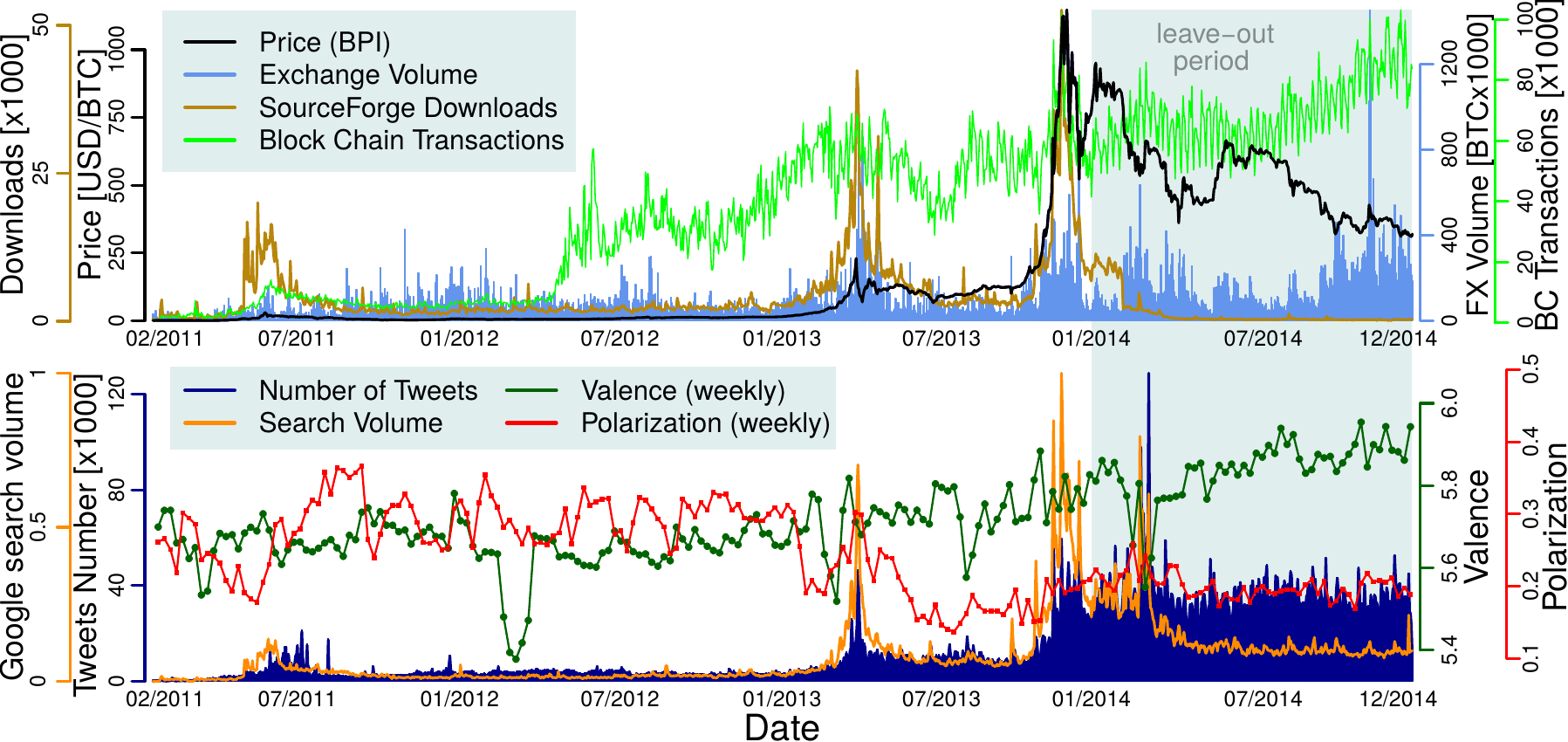}	
\caption{\textbf{Time series of data volumes in the Bitcoin ecosystem.} Interactive version: \texttt{www.sg.ethz.ch/btc}}
  \label{fig:DataTS}
\end{figure}

\paragraph{Bitcoin social economic and signals} We apply our approach to the case of
trading Bitcoin based on social and economic   signals of the Bitcoin ecosystem. We
set up a system that monitors different data sources, retrieving data in real
time in combination with historical time series. The data volumes recorded
during our study period of almost four years is shown in Figure
\ref{fig:DataTS}, and can be interactively browsed in our online
visualization\footnote{\texttt{www.sg.ethz.ch/btc}}. The signals we
measure, explained more in detail in the Materials and methods section include
economic signals  of price $P(t)$ and  returns $Ret(t)$, trading volume in a
wide range of Bitcoin exchange markets $FX_{Vol}(t)$. Furthermore, we
measure the economic signal of  transaction volume in the Block Chain
$BC_{Tra}(t)$, which measures the volume of usage of Bitcoin as a currency,
and the amount of downloads of the most important Bitcoin client $Dwn(t)$ as a
measure of growth in adoption of the Bitcoin technology. The social signals we
measure are the level of search volume in Google for the term "bitcoin"
$S(t)$, the word-of-mouth level as measured by the amount of \emph{tweets}
containing Bitcoin-related terms $T_{N}(t)$, and the emotional valence
$T_{Val}(t)$ and opinion polarization $T_{Pol}(t)$ expressed in those tweets
using lexicon-based approaches from psycholinguistics
\cite{Pennebaker2007,Warriner2013} (more details in Materials and methods).
All these signals are shown in Figure \ref{fig:DataTS},  illustrating the
large oscillations of price and other signals related to Bitcoin.

\section{Results}

\subsection{Data-driven Bitcoin trading strategy design}

\begin{figure*}[ht] \includegraphics[width=0.98\textwidth]{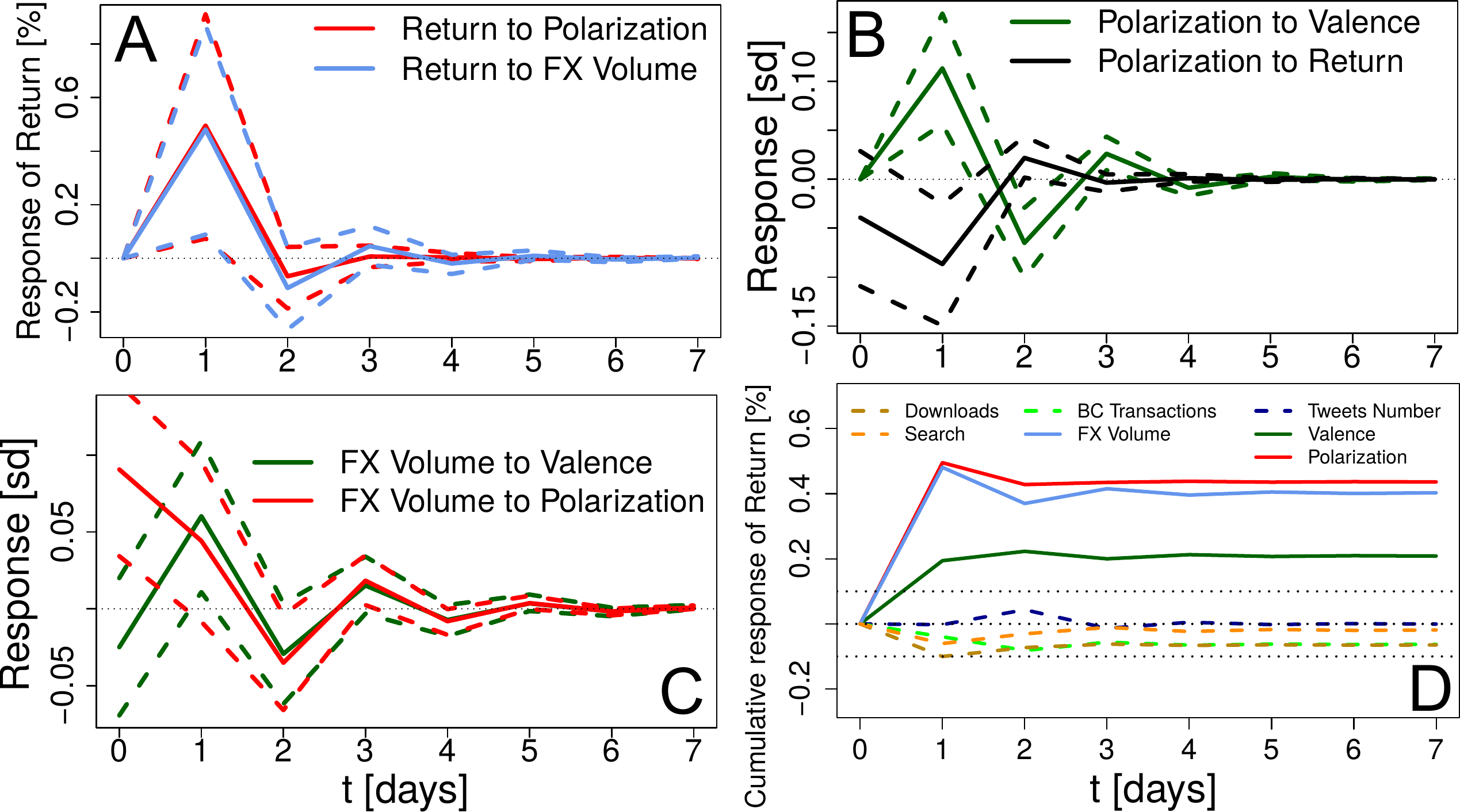}\\
\caption{\textbf{Results of Impulse Response Function analysis.} (A) Impulse Response Functions  of return to shocks in   Twitter
polarization and exchange volume, (B) of Twitter polarization to shocks in
return and Twitter valence,  and (C) of exchange volume to shocks in Twitter
valence and polarization (right). Solid lines show responses, dashed lines
show 95\% confidence intervals. (D) Cumulative Impulse Response Functions of
price return to changes in the other signals. Dashed lines indicate responses
below the 0.1\% level. } \label{fig:SelectedIRF} \end{figure*}

For our statistical analysis, we include all the data up to January 1st, 2014,
covering almost 3 years. After applying stationarity tests, we conclude that
the time series of price returns $Ret(t)$ can be assumed to be stationary, as
well as the first differences of the other seven signals (details on the
stationarity test results can be browsed in \texttt{www.sg.ethz.ch/btc} and
in the SI). As a consequence, we define our variable vector as:
\begin{eqnarray}
V(t) = [  Ret(t), \Delta FX_{Vol}(t), \Delta BC_{Tra}(t), \Delta Dwn(t),&\nonumber\\ 
\Delta S(t),  \Delta T_N(t), \Delta T_{Val}(t), \Delta T_{Pol}(t)]& \nonumber
\end{eqnarray}
composing the input to the multivariate analysis of our framework. We fit a
VAR as explained in Materials and methods over the analysis
period. We compute IRF for all pairs of variables, all results including VAR
estimates and IRF values can be browsed in \texttt{www.sg.ethz.ch/btc} and
in the SI. Here, we comment on the most relevant results, which serve as input
for our trading strategy design.

Figure \ref{fig:SelectedIRF} A shows the  IRF  of returns to shocks in
polarization and volume in exchange markets, where the response is measured in
return percentages.  Both polarization and exchange volume have significantly
positive influences in price returns one day after the shock, decreasing
rapidly afterwards. The increase of returns with polarization is consistent
with the hypothesis that disagreement fuels trading in speculative scenarios
\cite{Harris1993,Antweiler2004}, where information asymmetries fuel price
bubbles. Exchange volume also increases with polarization, as shown in Figure
\ref{fig:SelectedIRF} C, but the the relationship is instantaneous rather than
lagged as in the case of returns.

Figure \ref{fig:SelectedIRF} B shows the response of polarization in Twitter
to shocks in  returns and valence. The negative effect of polarization shows
that price drops lead to increases in polarization, signaling the disagreement
in the Bitcoin community due to price crashes. The pattern linking valence to
polarization is relevant, revealing that periods with increasing positivity in
expression precede stages of higher polarization. The role of valence can
further  be observed in the IRF of exchange volumes in  Figure
\ref{fig:SelectedIRF} C, in which valence has a significant effect.   The
combination of patterns of increasing polarization and exchange volume
following stages of increasing valence show the  relevance of valence in price
returns, in addition to the effects of polarization and exchange volume.

We further validated these results in two ways. First, we fit a VAR with lags
longer than a day, selecting the optimal lag that optimizes the Bayesian
Information Criterion. We found that a lag of 2 is optimal, but the results of
the fits and IRF analysis did not qualitatively change (see SI). Second, we
performed a Monte Carlo test, computing the impulse response functions for
time series with randomized permutations of the values. The results of these
permutation tests show are consistent with the above results, as reported in
the SI, showing the robustness of our approach.

\paragraph{Turning analysis into strategy} 

We summarize the above findings as stylized facts that can drive the decisions
of an algorithmic trader. We focus closer on the role of each signal into
returns, by computing the cumulative changes  given by the IRF analysis. This
way, we can identify which signals show a sizable pattern that precedes
changes in returns,  and filter out those that are not significant or can be
explained as confounds of the others. Figure \ref{fig:SelectedIRF} D shows the
results, measuring the cumulative change in return percentage when each one of
the other signals receives a shock of size one standard deviation. The three
signals with effects above the 0.1\% level are polarization, valence, and
exchange volume, reaching effects up to 0.5\% in one day that prevail through
time. Note that this is a relatively large value,  because trading results in
multiplicative returns. Such effect sizes have strong potential impact on the
profitability of trading strategies over long time periods. This allows us to
discard the rest of the signals, feeding into our trading strategy design by
producing four strategies: three strategies of positive sign, \emph{Valence},
\emph{Polarization}, and \emph{FXVolume}, and a fourth \emph{Combined}
strategy determined by a voting mechanism as explained in the Trading strategy
framework section.

\subsection{Bitcoin strategy evaluation}

To evaluate the profitability of our four strategies, we set up a benchmark
against random strategies and technical strategies, using the actual exchange
rate of BTC for USD in \texttt{bitfinex.com} as well as the Bitcoin Price
Index (see \texttt{www.sg.ethz.ch/btc} for results with BPI). Random
strategies sample a random number with $0$ mean at every time $t$, and
formulate a prediction based on the sign of the random number. Among technical
strategies, the simplest is \emph{Buy and Hold}, which simply buys BTC with
the initial capital at time $t=1$, selling it only once at the time when
profits are evaluated. The technical strategies we use are a benchmark of
simple standard predictions \cite{Biondo2013}: i) the \emph{Momentum}
strategy, which predicts that price changes at time $t+1$ will be the same as
at time $t$, ii) the Up and Down Persistency strategy \emph{UPD}, which
predicts that price increases at time $t$ are followed by decreases at time
$t+1$, and vice versa,  and iii) the Relative Strength Index strategy
\emph{RSI}, which  computes an additional time series of ratios of return sign
frequencies over a rolling window of five days, and predicts price changes
based on reversals of this time series (more details in \cite{Biondo2013}).

\begin{figure}[t]
\includegraphics[width=\textwidth]{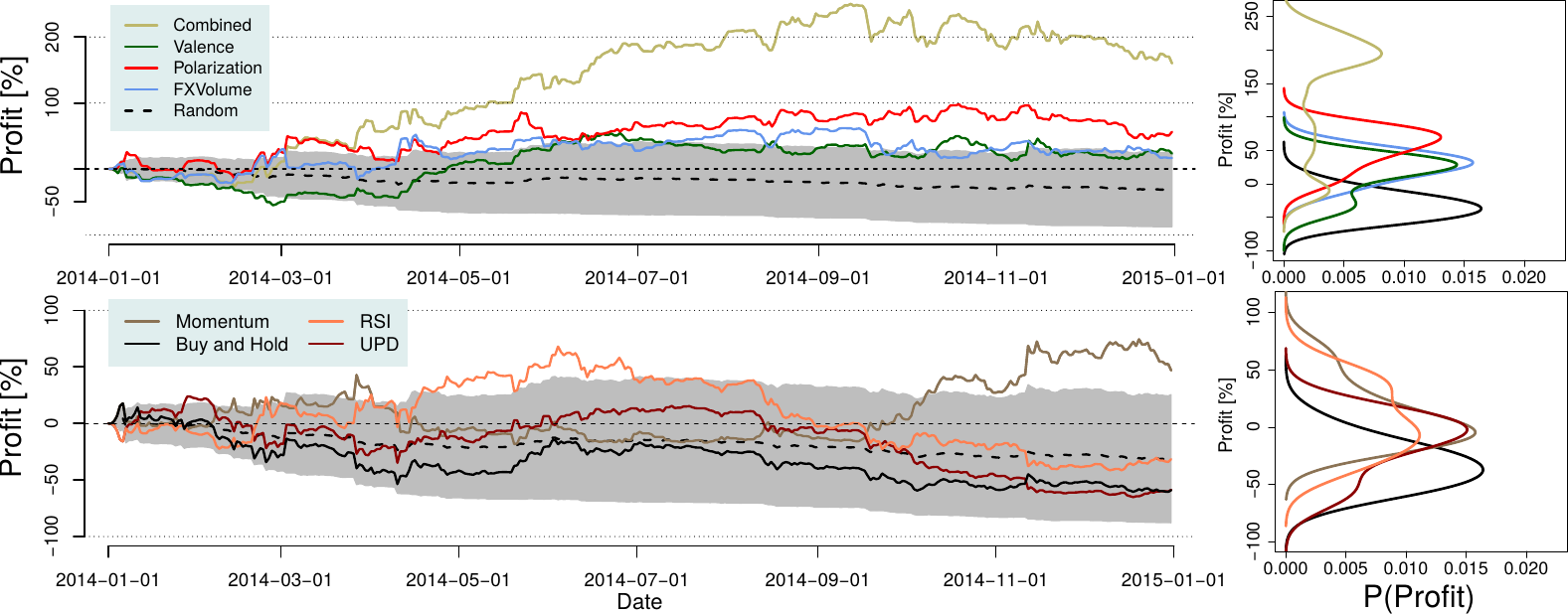}
\caption{\textbf{Profits of trading strategies.} Left: Time series of profit for our strategies (top), and technical strategies (bottom). Shaded areas show one standard deviation of the random strategy. Interactive version: \texttt{www.sg.ethz.ch/btc} Right: Kernel density plots of the profit of each strategy (bandwidth=15\%).}
  \label{fig:ReturnsTS}
\end{figure}

The simulation of each strategy produces a time series of profits
\begin{equation}
\text{Profit}(t) = \frac{C(t) - C(0)}{C(0)} * 100
\end{equation}
where $C(t)$ is the capital of the trader at time $t$ and $C(0)$ is the
initial investment capital. Figure \ref{fig:ReturnsTS} shows the time series
of profits for our four strategies and the technical strategies. In addition,
we compute the profit of \emph{Buy and Hold},  and the  results of the
simulation of 10000 random traders. The \emph{Valence}, \emph{Polarization}
and \emph{Combined} strategies clearly perform better than a random trader,
while the \emph{FXVolume} is not very far from the result of random traders.
Among the technical strategies, only \emph{RSI} and \emph{Momentum} are able
to eventually reach beyond the outcome of random traders, but are still
clearly outperformed  by the \emph{Polarization} and \emph{Combined}
strategies.

The stopping time of the simulation of a trading strategy is given by our
data, but a variety of factors might trigger a trader to stop trading earlier
in a real scenario \cite{Treleaven2013}. For that reason, we explore the
distribution of profits of each strategy, assuming that the trading stops at
any arbitrary point of our backtesting period. Thus, for each strategy we have
a set of profit values, one for each possible trading end date. The right panel
of Figure \ref{fig:ReturnsTS} shows the Kernel Density Plots of the
distributions of profits for each strategy. It can be appreciated that the
most profitable strategy is \emph{Combined}, followed by \emph{Polarization}
and then \emph{Valence} and \emph{RSI}. We quantitatively assessed this
result, through  Wilcoxon tests \cite{Wilcoxon1945} over the distributions of
profits (more details in SI), confirming the observation that the most
profitable strategies are \emph{Combined}, and \emph{Polarization}. More
precisely, the \emph{Combined} strategy gives profits beyond 100\% for most of
the time during the trading period.

While surveying cumulative returns is illustrative of the performance of the
strategies, the multiplicative nature of cumulative returns overweights early
positions and is biased towards the beginning of the evaluation period. To
properly evaluate trading strategies, we calculated the Sharpe Ratio
\cite{Sharpe1970}, measuring risk-corrected profits as:  $SR = \frac{\mu_R -
R_f}{\sigma_R}$, where $\mu_R$ and $\sigma_R$ are the mean and standard
deviation of the daily  rate of return of a strategy $R(t) = (C(t) -
C(t-1))/C(t-1)$.  $R_f$ is the "risk- free" return rate of a theoretical
investment that would give certain profit under no risk at all, which is often
estimated as the interest rate of high-quality sovereign bonds. At the time of
writing, some European sovereign bonds are giving interest rates close to zero
or even negative \cite{Edwards2014}, which motivates our conservative choice
of $R_f=0$.  The value of $SR$ is calculated in annualized units, taking into
account that Bitcoin can be traded 365 days a year.

\begin{table}[ht]
\centering
\begin{tabular}{lrrrrrrrrrr}
  \hline
& Combined & Polarization & Valence & FXVolume & Buy and hold \\ 
\hline
$SR$ & 1.7653  & 1.0120 & 0.6410 & 0.5738 & -0.7741\\ 
$\mu_R$ & 0.3229 & 0.1779 & 0.1183 & 0.1082 & -0.1635 \\ 
\hline
& Momentum & UPD & RSI & DJIA & Random \\ 
\hline
$SR$ & 0.9146 & -0.8990 & -0.1772 & 0.7995 & -1.6590  \\ 
$\mu_R$ & 0.1625 & -0.1736 & -0.0346 & 0.0345 & -0.0963 \\ 
\hline

\end{tabular}
\caption{\label{tab:Sharpe}Sharpe Ratios and mean daily returns of strategies.} 
\end{table}

Table \ref{tab:Sharpe} reports the Sharpe Ratio $SR$ and the mean daily return
$\mu_R$ for all strategies, as well as for the  \emph{DJIA} and the average of
10000 random traders.  The Sharpe Ratio analysis is consistent with  the
results of the cumulative returns analysis, showing that the \emph{Combined}
strategy provides the highest returns, with the best $SR$ value above 1.75 and
with daily returns above 0.3 \% per day. The profitability of these strategies
illustrate how social media sentiment can produce positive returns on
investment, especially when including polarization measures beyond the trivial
quantification of valence or mood.

\subsection{Costs and risks of the Combined strategy}

\begin{figure}[h]
\centering
\includegraphics[width=0.48\textwidth]{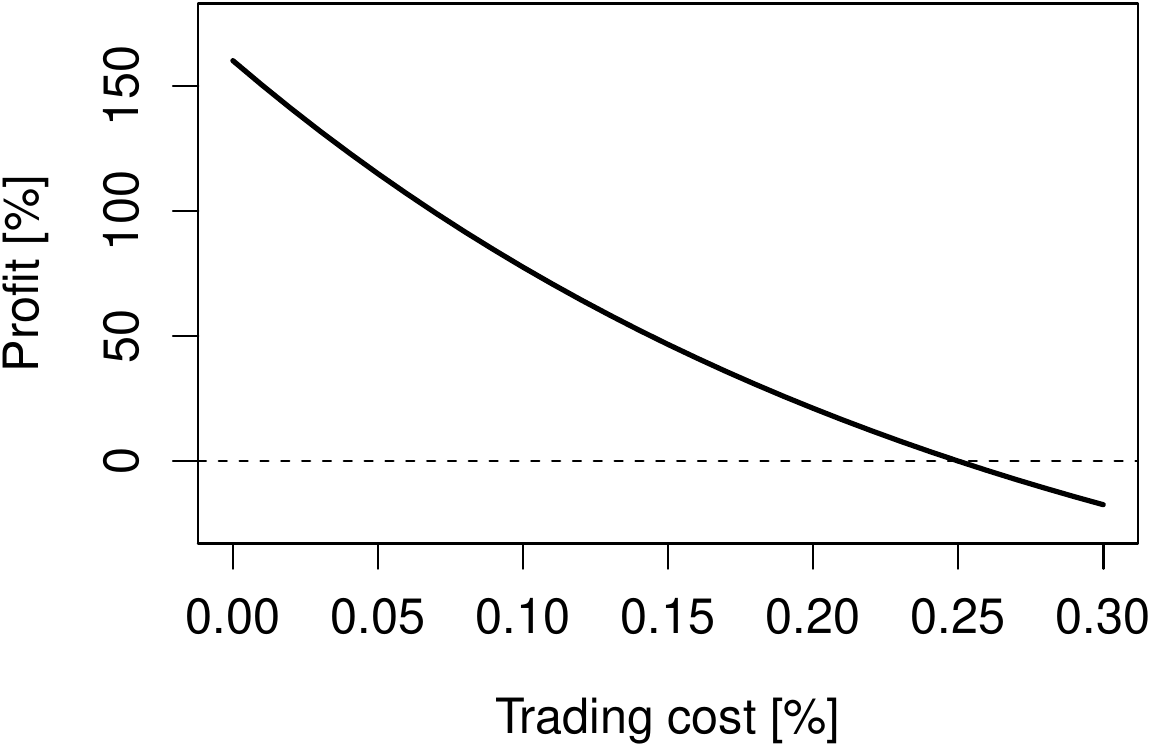}
\caption{\textbf{Final profit of the \emph{Combined} strategy versus trading cost.}
  \label{fig:CostPlot}}
\end{figure}

To understand better the possible weaknesses of the \emph{Combined} strategy, we run
a series of tests to evaluate the role of trading costs and additional risks.
Trading Bitcoin in an online market usually comes at a cost, which often
depends on the activity and the traded capital. These trading costs should not
be confused with the transaction fees in the Block Chain \cite{Nakamoto2008},
which do not depend on the transacted cost and are not associated to any
market of exchange to other currencies.  Trading costs can potentially erode
the profitability of trading strategies,  especially if they require many
movements. We simulated the same backtests for  costs increasing from 0 to
0.3\% of the exchanged capital, a value well above the maximum costs of major
trading platforms \cite{Bitfinex2014}. As a simplification, we assume that
buying, selling, and borrowing costs are the same, yet their values might
depend on the trading volume of a strategy \cite{Bitfinex2014}. Figure
\ref{fig:CostPlot} shows the final profits of the \emph{Combined} strategy,
which  decrease monotonically with trading costs. The strategy is still
highly profitable for low costs, but for costs above 0.25\%, the strategy is
not profitable any more. Furthermore, we repeat this analysis assuming the
limitation that daily positions need to be forcefully closed at the end of each
trading period (shown in SI), finding a decrease in returns but that the
strategy is still profitable for trading costs of 0.1\%, a typically high cost of
current exchange platforms \cite{Bitfinex2015}.

In this application, the leave-out period is characterized by decreasing BTC
prices. Thus, it is important to evaluate the role of each possible trading
action: longing when BTC are bought to be sold later, and shorting when they
are borrowed and traded as explained above. We repeat the backtesting of the
\emph{Combined} strategy allowing only short and only long positions,
following the methodology of \cite{Preis2013}. As reported more in detail in
the SI, the only short strategy yields higher cumulative returns than the only
long strategy, as expected from a period in which prices decrease steadily.

We test further properties of the behavior of the \emph{Combined} strategy
in the leave-out period. The distribution of daily returns of the
\emph{Combined} strategy during the leave-out period follows a lognormal
distribution, as tested through maximum likelihood fits and Kolmogorov-Smirnov
tests (see SI). The time series of returns of this strategy is also not
autocorrelated and can be considered stationary (see SI for stationarity tests
of daily returns). This additional analysis shows that the  high profitability
of the \emph{Combined} strategy is not due to risky correlations in the
behavior of the trading strategy.

\section{Concluding remarks}

Our work applies established methods of time series analysis and computational
finance to integrate the analysis, design, and evaluation of trading
strategies and social and economic signals. We have shown that our approach
successfully reveals temporal patterns in the Bitcoin ecosystem, in particular
the relation between price returns and the signals of exchange volume and
Twitter valence and polarization.  Our statistical analysis is robust to noise
correlations and the finite nature of time series, providing a consistent set
of results that we can apply to strategy design. We evaluated the
profitability of our strategies through data-driven simulations of a
computational model of a trader, showing that a strategy that combines
valence, polarization, and exchange volume can reach very high profits in less
than a year. The added value of including polarization in our analysis
constitutes evidence that  collective factors of emotions and opinions have
the potential to predict financial returns,  beyond trivial macroscopic
aggregates like average valence.

Our framework can be applied to other trading scenarios in which social
signals are available, like in the case of  company stock trading driven by
sales data, news information, and social media sentiment towards a company.
The general nature of our methods are of special relevance for real trading
scenarios, as the stylized facts we use to design strategies provide a
tractable explanation for their mechanisms. This allows traders to understand
and evaluate the principles of the algorithmic trading strategies designed in
our framework. Such tractability is an advantage in comparison to more
complex, non-linear, or subsymbolic models that do not have straightforward
interpretations. Nevertheless, improvements can be expected from the addition
of longer time lags, higher frequency trading, and real-time optimization
approaches. Furthermore, the rules that drive our trading strategies do not
require retraining or calibration during trading, and the social and economic
signals we employ can be quantified during a day in order to have an instant
trading decision ready at the beginning of the next day. Our application to
Bitcoin trading is thus realistic, making use of shorting options and
performing well under the typical trading costs of Bitcoin markets
\cite{Bitfinex2014}.

The application of our results should be taken with caution.  Historical
profit through backtesting do not necessarily predict future ones,  and the
information sources analyzed here could be adopted by Bitcoin traders. Our
evaluation goes as far as the representativity of the leave-out sample, and
future research should evaluate the performance of our approach when prices
rise and when traders are aware of the existence of our trading strategies.
Financial markets are known to quickly absorb knowledge, as it happened with
the inclusion of search trends data in stock trading \cite{Curme2014}. It is
also difficult to estimate the scalability of automatic trading strategies, as
financial markets are complex adaptive systems that react to trades of large
volume. Furthermore, systemic risk emerges from algorithmic trading, creating
\emph{flash crashes} due to algorithmic resonance \cite{Cusumano2014}. In
addition, structural changes and additional risks in borrowing and lending
Bitcoin for shorting can emerge when  exchange markets close or governments
regulate Bitcoin, changing the rules of the game in a way such that our
trading strategies  might not work any more.

With our study, we have shown  that it is possible to turn social signals into
profit. This extends the range of  typical business applications for social
media data like viral marketing or user engagement. Specifically, our
combination of statistical analysis and backtesting serves as a framework for
future applications of social media data in algorithmic trading. It allows a
robust validation of strategy profits and a clear understanding of the system
dynamics behind these profits.  The application of our framework to Bitcoin
trading illustrates that (asymmetric) information and profit are two
manifestations of the same thing, and how traders can apply these macroscopic
information sources to derive large profits. We foresee that the applications
of social signals to finance will reach far beyond Bitcoin, not only to make
private profit but also to understand the dynamics of individual and
collective decisions and emotions.

\section{Materials and methods}

\paragraph{Stationarity tests.}  Before fitting the VAR model, we test  the
stationarity of each time series through two alternative tests: i) the
Augmented Dickey Fuller (ADF) test \cite{Fuller2009}, which has the null
hypothesis that the tested time series is \emph{non-stationary}, and ii) the
Kwiatkowski-Phillips-Schmidt-Shin (KPSS) test \cite{Kwiatkowski1992}, with the
null hypothesis that the time series is \emph{stationary}. Under these two
tests, it can be considered safe that a time series is stationary if it passes
the ADF test  with  a p-value  below $0.05$ and does not pass the KPSS test,
giving a p-value above $0.1$ \cite{Lutkepohl2007,Garcia2014}. We first analyze
the time series of levels of each signal  $X(t)$, applying the differentiation
operator $\Delta X(t) = X(t) - X(t-1)$ until each time series is stationary.
This step is inspired in the Box-Jenkins method of ARIMA time series analysis
\cite{Anderson1976}, and it is usual to reach stationarity after first
differences \cite{Lutkepohl2007,Garcia2014}. The stationary properties of
these time series imply that their means and standard deviations are bound,
allowing us to renormalize them through the Z-transformation $Z(t) = (X(t) -
\mu_X)/\sigma_X$, where $\mu_X$ and $\sigma_X$ are the mean and standard
deviation of each time series. This way, all time series have the same scale
and variance, and their effects in statistical analysis can be compared.

\paragraph{Impulse response function analysis.} In the impulse analysis, we
correct for the correlations in $\epsilon$ in two ways. First, we apply
orthogonalized impulses of unit covariance, creating a shock of one standard
deviation in a variable under the error correlations of the VAR
\cite{Zeileis2004}. Second, we apply bootstrapping on the resulting responses
by producing surrogate time series from resampling the residuals
\cite{Lutkepohl2007}. This way, we numerically compute confidence intervals of
the responses in a very strict way, avoiding false positives and taking into
account the finite size of the analysis period.   In our case, we create
$10,000$ bootstrap samples to estimate 95\% confidence intervals of the
responses. As a result, we simultaneously measure the dynamics of the system
and test their statistical significance.

\paragraph{Trading based on predictions.} During each timestep, the prediction
function makes a forecast either based on Equation \ref{eq:prediction}, or
based on the price time series for technical strategies. Positive predictions
translate into \emph{buy} decisions when the trader does not own the asset,
and \emph{hold} if it does.  When the predictor takes value 0, no change is
done and the previous position is imitated.  Negative predictions translate
into \emph{sell} positions when the trader owns the asset or \emph{short} when
it does not own it. Shorting works as follows: Traders can make profit from
correct predictions of price drops even if they do not own the asset predicted
to drop in price. This is implemented by borrowing the asset, selling it first
and buying it later for a lower price. The limitation for borrowing is usually
imposed on the amount of capital already held by the trader, and often incurs
in additional trading costs and legal regulations \cite{Chance2015}. The
simulation of each strategy produces a time series of profits, allowing us to
measure their profitability based on historical data.

Buy and sell orders have respective costs $c_b$ and $c_s$, which are
proportional to the total traded capital. In our case we assume all  costs are
equal $c=c_b=c_s$, leaving particular realizations of the costs as open for
future research. We compute daily cumulative returns when trading stops at
t+1, holding USD or selling BTC at the price of t+1.   Our trading simulations
have a limit on short selling set by the amount of capital held by the trader
and assume that short selling needs to be instantly executed, i.e. short
positions are limited to one iteration. In summary, the strategy we execute is
a   single-asset backtesting scenario in which 100\% of the capital is
invested at each time step and shorting is limited. The pseudocode of this
simulation is shown in Algorithm \ref{alg}.

\begin{algorithm}[h]
 nUSD = 1; CR[1]=1\;
 \For{each t from 1 to T-1}{
  \uIf{prediction(X,t) == 1 and nBTC == 0}{
  		nBTC = nUSD * (1-$c_b$) / P[t]	\;
		nUSD = 0 
   }
   \uElseIf{prediction(X,t) == -1 and nBTC > 0}{
		nUSD= nBTC * (1-$c_s$) * P[t]	\;
		nBTC = 0  						
   }
   \uElseIf{prediction(X,t) == -1 and nBTC == 0}{
   		nBTCb = nUSD / P[t] \;
        nUSD = nUSD + nBTCb * (1-$c_s$) * P[t] -  nBTCb * P[t+1] / (1-$c_b$)
   }
   CR[t+1] = nUSD + nBTC * P[t+1] *(1-$c_s$)\;

 }
 \caption{Algorithm of trading simulation. \label{alg}}
\end{algorithm}

\paragraph{Economic signals from financial data} The establishment and
bankruptcy of various Bitcoin exchange markets motivated the creation of the
Bitcoin Price Index (BPI) \cite{Coindesk2013}. The BPI combines a set of price
indices from well-performing exchange marketplaces to provide  a reference for
BTC/USD exchange rates, and is accepted as a standard measure of Bitcoin price
in economics \cite{Brito2014,Shadab2014,Kristoufek2015}. We use the daily closing prices of
each day $t$ at 23:59 GMT from \texttt{coindesk.com}, composing the time
series  of price $P(t)$ from February 1st, 2011 to December 31th, 2014, shown
in the top panel of Figure \ref{fig:DataTS}. The BPI is not necessarily
tradeable, and for that reason we evaluate our trading strategies with the
actual exchange rate of BTC for USD in \texttt{bitfinex.com}, one of the
largest markets reported in \texttt{coindesk.com}. We also retrieved the
daily volume of BTC exchanged in 80 online markets for other currencies from
\texttt{bitcoincharts.com}. Aggregating all these data sources, we compose an
Internet-wide measurement of Foreign eXchange (FX) volume of BTC traded every
day $FX_{Vol}(t)$,  including more than 152 Million BTC in exchange trades as
we recorded in early 2015.

Every purchase of products and services in BTC leaves a trace in the
\emph{Block Chain}, the distributed ledger that records all transactions in
the Bitcoin network. We construct a time series with the daily amount of Block
Chain transactions $BC_{Tra}(t)$, as measured by \texttt{blockchain.info} every
day at 18:15:05 UTC, which we approximate to 00:00 GMT of the next day.  While
some data is lost in this additional delay of few hours, further research can
provide  more precise measurements up to the minute level using the raw
information in the \emph{Block Chain} itself as in \cite{Garcia2014}. This
way, we include more than 55 million transactions in the studied period,
measuring  the overall activity of the system when using Bitcoin as means of
exchange. In addition, we measure the growth of the Bitcoin market through its
amount of adopters, using the operationalization of measuring the amount of
downloads of the most popular Bitcoin
client\footnote{\texttt{http://sourceforge.net/projects/bitcoin}} \cite{Garcia2014},
daily binned in line with other time series. The resulting time series of
downloads $Dwn(t)$ is shown in the top panel of Figure \ref{fig:DataTS}.

\paragraph{Social signals} We record the overall interest towards Bitcoin
through information search,  as quantified by the Google trends volume for the
term "bitcoin", $S(t)$, as recorded in early 2015 and shown in the bottom
panel of Figure \ref{fig:DataTS}. We choose the search term "bitcoin" instead
of the "Bitcon - Currency" topic, which was introduced as a functionality of
Google Trends during our analysis period.  While the topic approach can be
more precise for demographics and motive analysis \cite{Yelowitz2015}, we
follow a homogeneous approach including only the term trend data that was
available during the whole study period. It is important to note that Google
Trends data is provided with an additional lag of one day and on the basis of
Pacific Standard Time instead of GMT, adding almost another day of lag. While
this is not an issue for the historical analysis, the evaluation of any
trading strategy using $S(t)$ needs to take into account this additional
delay.

We track the attention of social media about Bitcoin in Twitter via  the Topsy
data service\footnote{\texttt{http://topsy.com/}} . From the full track of data
accessible by Topsy \cite{Aitken2013}, we focus on tweets that contain Bitcoin
terms as in previous research \cite{Garcia2014}, finding a total of
$19,578,671$ Bitcoin-related  tweets. The first social signal we extract from
Twitter  is the daily amount of unique tweets about Bitcoin $T_{N}(t)$ binned
in 24 hour windows starting at 00:00 GMT, measuring the level of word-of-mouth
and attention towards Bitcoin and shown in the bottom panel of Figure
\ref{fig:DataTS}. We continue by measuring the collective emotional valence
with respect to Bitcoin, as expressed through the text of Bitcoin-related
tweets. Valence is considered the most important dimension of affect,
quantifying the degree of pleasure or displeasure of an emotional experience
\cite{Russell2003}. The expression of valence through text is a common
practice in psychological research, in which lexicon techniques are used to
empirically measure emotions \cite{Bradley1999,Garcia2012}. We measure the
average daily valence of Bitcoin-related tweets through  a state-of-the-art
lexicon technique \cite{Warriner2013},  which improves the previous ANEW
lexicon method \cite{Bradley1999} with more than $13000$ valence-coded words.
We  compute the daily average Twitter valence about Bitcoin during day $t$ in
two steps: First we measure the frequency of each term in the lexicon during
that day, and second we compute the average valence weighting each word by its
frequency. This measurement matches more than 50 million valence-carrying
tokens, and produces the time series of Twitter valence $T_{Val}(t)$.

Our last social signal is opinion polarization, which builds up on measuring
the semantic orientation of words into positive and negative evaluation terms
\cite{Osgood1964}. We apply the LIWC psycholinguistics lexicon-based method
\cite{Pennebaker2007}, and expand its lexicon of stems into words by matching
them against the most frequent English words of the Google Books dataset
\cite{Lin2012}. As a result, we consider 3463 positive and 4061 negative terms
that appear as more than 8 million Twitter tokens. We compute the daily
polarization of opinions  in Twitter around the Bitcoin topic $T_{Pol}(t)$,
calculating the geometric mean of the daily ratios of positive and negative
words per Bitcoin-related tweet.  Note that, instead of repeating a
measurement of valence through two different lexica, we quantify polarization
as a complementary dimension to emotional valence. This way, opinion
polarization measures the simultaneous coexistence of positive and negative
subjective content, rather than its overall orientation
\cite{Osgood1964,Tumarkin2001}.


\noindent \textbf{Acknowledgements:} We thank Emre Sarigol for his technical assistance.\\
\noindent \textbf{Funding:} This work was funded by the Swiss National Science Foundation (CR21I1\_146499). \\
\noindent \textbf{Authors' contributions:} DG conceived the research, retrieved data, and performed the analyses. DG and FS wrote the manuscript.\\
\noindent \textbf{Competing interests:} The authors declare that they have no competing financial interests.\\
\noindent \textbf{Ethics statement:} This research is based on observational data shared publicly. No personal or individual information was retrieved, stored, or analyzed.\\
\noindent \textbf{Data and materials availability:} All data used for this article is publicly available through the corresponding Application Programming Interfaces. Processed data and scripts to reproduce these results are available at \texttt{https://www.sg.ethz.ch/btc}

\end{document}